\pdfoutput=1
\documentclass[fleqn,10pt]{wlscirep}
\usepackage[T1]{fontenc}
\usepackage{bm,graphicx,color}

\newcommand{\cc}[1]{{\color{green} \bf #1}}
\newcommand{\del}[1]{}

 \newcommand{\var}[1]{}

\title{Plasmonic antennas with electric, magnetic, and electromagnetic hot spots based on Babinet's principle}
\author[1]{M.~Hrto\v{n}}
\author[2]{A.~Kone\v{c}n\'a}
\author[1]{M.~Hor\'ak}
\author[1,3]{T.~\v{S}ikola}
\author[1,3,*]{V.~K\v{r}\'apek}

\affil[1]{Central European Institute of Technology, Brno University of Technology, Purky\v{n}ova 123, 612 00 Brno, Czech Republic}
\affil[2]{Materials Physics Center CSIC-UPV/EHU, Paseo Manuel de Lardizabal 5, 20018 San Sebasti\'an, Spain}
\affil[3]{Institute of Physical Engineering, Brno University of Technology, Technick\'a 2, 616 69 Brno, Czech Republic}
\affil[*]{vlastimil.krapek@ceitec.vutbr.cz}


\begin{abstract}

We theoretically study plasmonic antennas featuring areas of extremely concentrated electric or magnetic field, known as hot spots. We combine two types of electric-magnetic complementarity to increase the degree of freedom for the design of the antennas: bow-tie and diabolo duality and Babinet's principle. We evaluate the figures of merit for different plasmon-enhanced optical spectroscopy methods: field enhancement, decay rate enhancement, and quality factor of the plasmon resonances. The role of Babinet's principle in interchanging electric and magnetic field hot spots and its consequences for practical antenna design are discussed. In particular, diabolo antennas exhibit slightly better performance than bow-ties in terms of larger field enhancement and larger Q factor. For specific resonance frequency, diabolo antennas are considerably smaller than bow-ties which makes them favourable for the integration into more complex devices but also makes their fabrication more demanding in terms of spatial resolution. Finally, we propose Babinet-type dimer antenna featuring electromagnetic hot spot with both the electric and magnetic field components treated on equal footing.

\end{abstract}

\begin{document}

\flushbottom
\maketitle
\thispagestyle{empty}

\section*{Introduction}

Plasmonic antennas are metallic particles widely studied for their ability to control, enhance, and concentrate electromagnetic field~\cite{Novotny2011}. Strikingly, the field in the vicinity of plasmonic antennas, so-called near field, can be focused to deeply subwavelength region. At the same time, the field is strongly enhanced (in comparison to the driving field, which can be e.g. plane wave). The mechanism behind the focusing of the field are localised surface plasmons (LSP) -- quantized oscillations of the free electron gas in the metal coupled to the evanescent electromagnetic wave propagating along the boundary of the metal.

In judiciously designed plasmonic antennas, local spots of particularly enhanced electric or magnetic field can be formed, referred to as hot spots. They typically arise from the interaction between adjacent parts a plasmonic antenna separated by a small gap~\cite{Nie1102,PhysRevLett.82.2590} but they can be also based on the lightning rod effect (a concentration of the field at sharp features of the antenna)~\cite{PhysRevB.64.045411,Gao:14,doi:10.1021/acsnano.7b07401} or combination of both. In various studies, electric hot spots have been reported over a broad spectral range from THz~\cite{Gao:14} (hot spot size $\lambda/60000$ predicted from electromagnetic simulations) to visible~\cite{Benz726} (hot spot size $\lambda/600$ and enhancement $>500$).

Depending on the enhanced field, hot spots can be classified as electric, magnetic, or electromagnetic. A variety of plasmonic antennas with specific shapes, sizes, and materials exists for both electric and magnetic hot spots. Electric hot spots have been observed in the nanorod dimer antennas, bow-tie antennas~\cite{Zhou:09}, or chains of plasmonic nanoparticles~\cite{Nie1102,doi:10.1021/ja2015179}. Magnetic hot spots are formed in diabolo antennas~\cite{doi:10.1021/nl103817f}, nanorings~\cite{doi:10.1021/nl400798s}, or split-ring resonators~\cite{doi:10.1093/jmicro/dfy010}. Electromagnetic hot spots with simultaneous enhancement of both electric and magnetic field are unique for plasmonic antennas~\cite{doi:10.1021/acsphotonics.6b00857}. Their formation has been observed in the dielectric resonators (silicon nanodimers)~\cite{doi:10.1021/acs.nanolett.5b00128}.

Hot spots can be involved in many application including surface enhanced Raman scattering~\cite{Nie1102,PhysRevB.95.115441,tian2018plasmonic}, improved photocatalysis~\cite{YANG2016612}, or fluorescence of individual molecules~\cite{Kinkhabwala2009}. Metallic resonators with enhanced magnetic field (magnetic hot spots) are regularly used to increase the efficiency of magnetic spectroscopies such as electron paramagnetic resonance~\cite{BLANK20131937}. Electromagnetic hot spots can be useful for studies of materials with combined electric and magnetic transitions such as rare earth ions~\cite{PhysRevLett.114.163903,PhysRevB.96.224430}. combined enhancement of electric and magnetic field finds applications also in optical trapping~\cite{Juan2011}, metamaterials~\cite{Smith788}, or non-linear optics~\cite{Klein502}. 

For experimental characterization of plasmonic hot spots, the available methods is scanning near-field optical microscopy~\cite{doi:10.1021/acs.nanolett.7b00503,Caselli2015,doi:10.1021/acs.nanolett.5b00128}, photon scanning tunneling microscope~\cite{PhysRevLett.82.2590}, or photothermal-induced resonance~\cite{doi:10.1021/nl401284m}.

{\it Bow-tie} geometry of plasmonic antennas features particularly strong electric hot spot. Bow-tie antennas are planar antennas consisting of two metallic triangular prisms (wings) whose adjacent apexes are separated by a subwavelength insulating gap. The hot spot arises from the interaction between the apexes combined with the lightning rod effect (the charge of LSP accumulates at the apexes). When the insulating gap is replaced with a conductive bridge, a {\it diabolo} plasmonic antenna is formed. Instead of charge accumulation, electric current is funneled through the bridge, resulting into magnetic hot spot. Both the bow-tie and diabolo antennas have been frequently studied.

Various optimization and modification approaches have been proposed with the aim to enhance the properties of the bow-tie and diabolo antennas, including the gap optimization~\cite{doi:10.1021/jz302018x}, fractal geometry~\cite{Cakmakyapan:14}, or Babinet's principle. Babinet's principle relates the optical response of a (direct) planar antenna and a complementary planar antenna with interchanged conductive and insulating parts. Both the direct and complementary antennas shall support LSP with identical energies, but with interchanged electric and magnetic near field~\cite{PhysRevLett.93.197401,doi:10.1021/nl402269h}. Consequently, when the direct antenna features electric hot spot, the complementary antenna features magnetic hot spot and vice versa. The validity of Babinet's principle for the plasmonic antennas has been experimentally verified~\cite{Bitzer:11,PhysRevB.76.033407}, although some quantitative limitations have been found in particular in the visible~\cite{Mizobata:17,Horak2018SciRep}.

A unique combination of Babinet's complementarity and bow-tie/diabolo duality extends a degree of freedom for the design of plasmonic antennas featuring hot spots. In our contribution we compare the two antennas featuring electric hot spot (bow-tie and complementary diabolo) and the other two featuring magnetic hot spot (diabolo and complementary bow-tie). By electromagnetic modeling we retrieve the characteristics of the hot spots and figures of merit of relevant plasmon-enhanced optical spectroscopy methods. Finally, we design Babinet dimer antennas featuring electromagnetic hot spots, rather unique in the field of plasmonics.

\section*{Results and discussion}

\begin{figure}[ph!]
  \begin{center}
    \includegraphics[clip,width=0.8\linewidth]{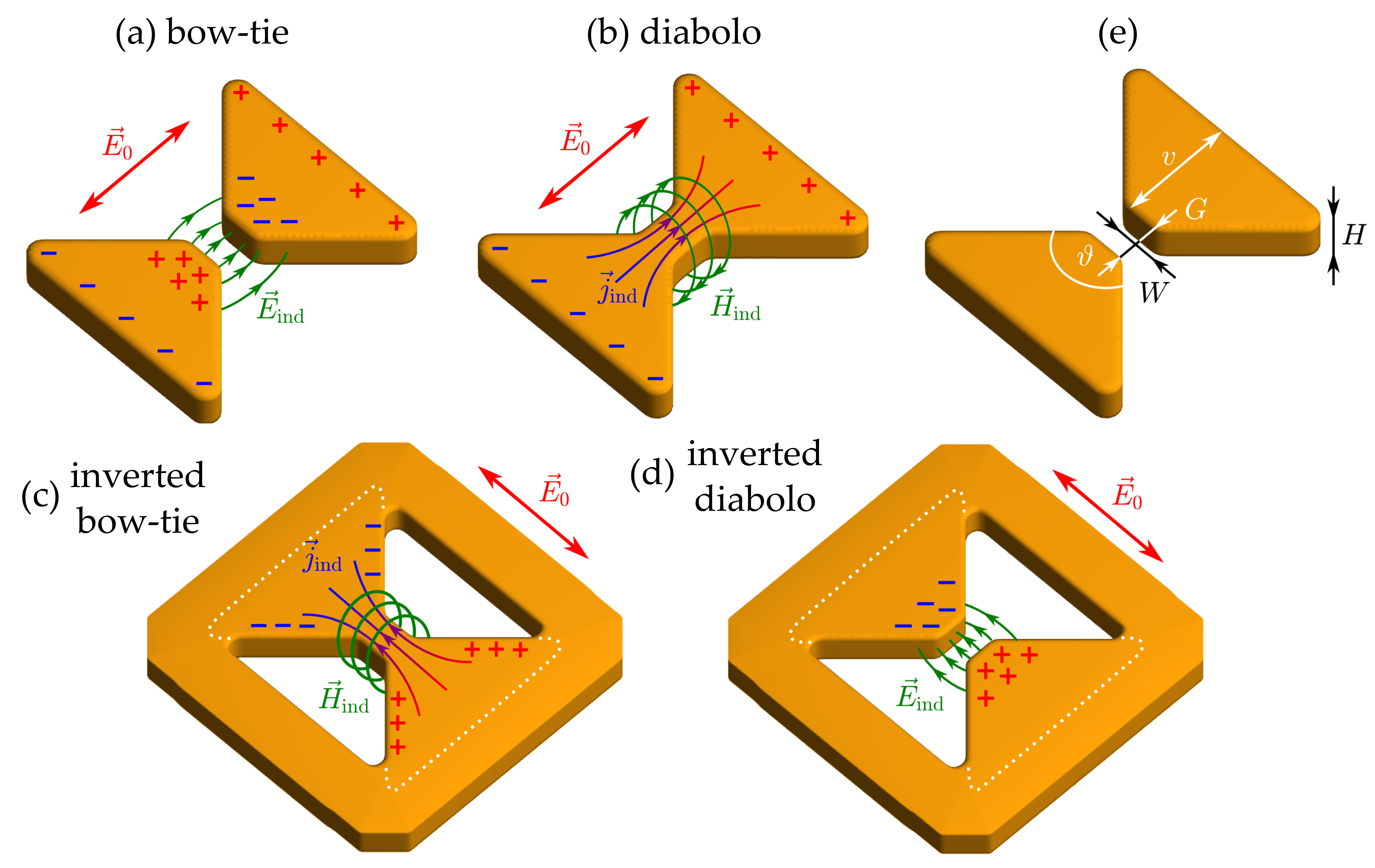}
    \caption{\label{figNO1} Schemes of four plasmonic antennas featuring hot spots: (a) Bow-tie, (b) diabolo, (c) inverted bow-tie, (d) inverted diabolo. Metallic and hollow parts are represented by yellow and white color, respectively. Driving electric field is indicated by red arrows. Charge or current accumulation and formation of the electric or magnetic hot spot are shown as well. Dotted white line in (c) indicates the qualitative correspondence between the diabolo and inverted bow-tie antennas, while dotted white line in (d) indicates a similar correspondence between the bow-tie and inverted diabolo antennas. Dimensions of the antennas are shown in (e).}
  \end{center}
\end{figure}

\subsection*{Plasmonic antennas, modes and hot spots}

Plasmonic antennas involved in the study and their operational principle are illustrated in Fig.~\ref{figNO1}. {\it Bow-tie} antenna consists of two disjoint triangular gold prisms. Oscillating electric field applied along the long axis of the antenna drives the oscillations of charge that is funneled by the wings of the antenna and accumulated at the adjacent tips [Fig.~\ref{figNO1}(a)]. Combined effects of plasmonic field confinement, charge funneling, and charge concentration (lightning rod effect) give rise to an exceptionally high field in the area between the triangles, by orders of magnitude higher than the driving field. In {\it diabolo} antenna, the triangles are connected with a conductive bridge, through which a concentrated current flows instead of charge accumulation [Fig.~\ref{figNO1}(b)]. A magnetic hot spot is formed around the bridge. {\it Inverted bow-tie} antenna is formed by two disjoint triangular apertures in otherwise continuous gold film. Babinet's principle predicts that for a complementary illumination (i.e., transverse oscillating electric field) a complementary magnetic hot spot is formed. This can be understood also intuitively as the antenna resembles a rotated diabolo antenna [see dotted line in Fig.~\ref{figNO1}(c)]. Finally, {\it inverted diabolo} antenna, which on the other hand resembles the bow-tie antenna, features electric hot spot Fig.~\ref{figNO1}(d).

The dimensions of the antennas are schematically depicted in Fig.~\ref{figNO1}(d). The thickness of the gold film is set to $H=30$\,nm. The size of the right isosceles triangles (i.e. $\vartheta=90^{\circ}$) is described by the wing length $v$. Opposite triangles share a common apex. The isolating gap in bow-tie antennas and the conductive bridge in diabolo antennas have the length $G$ equal to the width $W$. These dimensions do not scale with the size of the antenna (the only scalable parameter is thus $v$) and are set to 30\,nm to reflect common fabrication limits. In general, one could expect stronger hot spots for narrower gaps or bridges due to stronger charge or current concentration. All edges are rounded with a radius of 10\,nm. The antennas are situated on a semi-infinite glass substrate (refractive index 1.47). The dielectric function of gold is taken from Johnson and Christy.~\cite{PhysRevB.6.4370}.

\begin{figure}[h!]
  \begin{center}
    \includegraphics[width=0.8\linewidth]{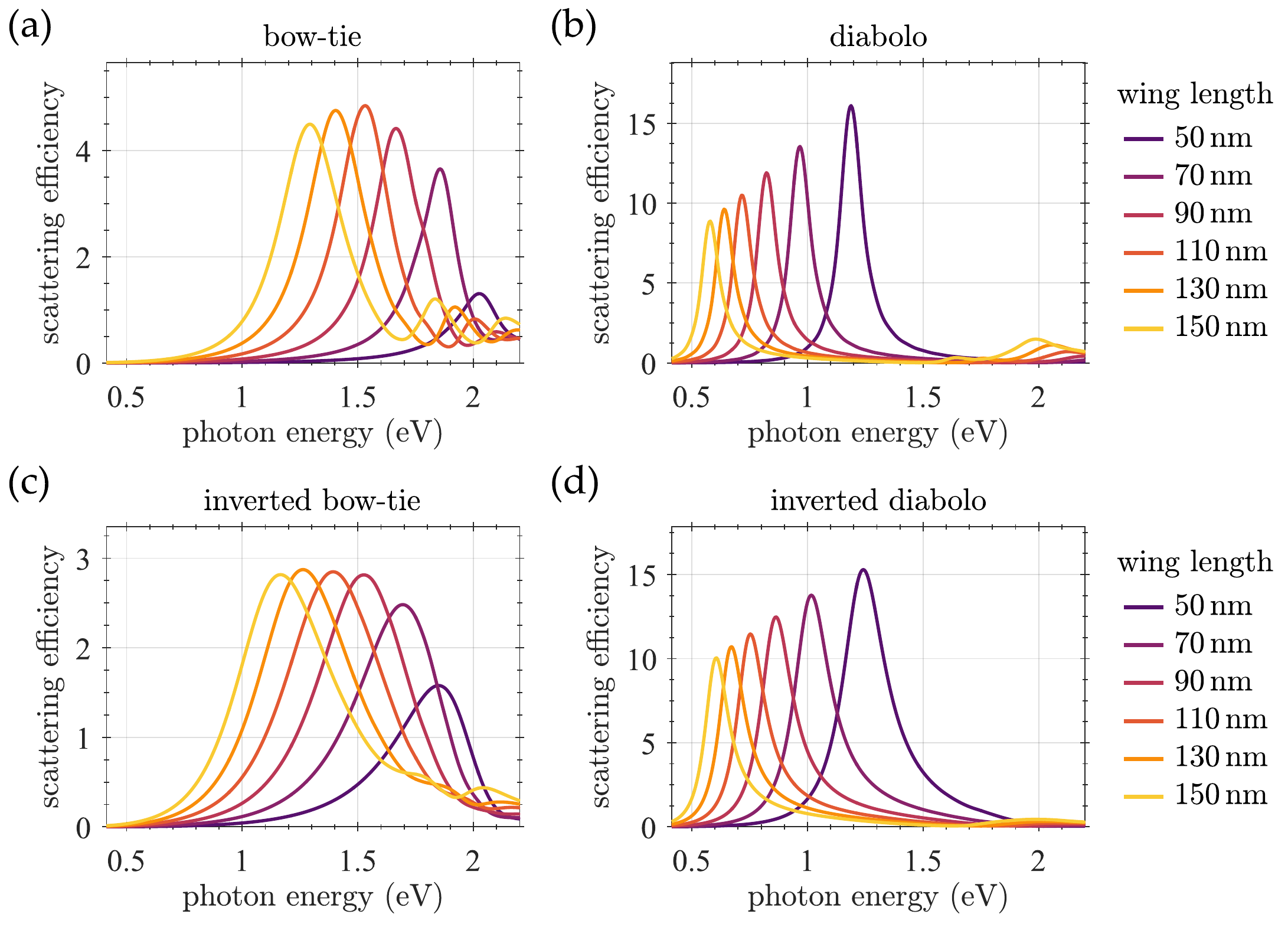}
    \caption{\label{figNO2} Spectral dependence of the scattering efficiency $Q_\mathrm{scat}$ of (a) bow-tie, (b) diabolo, (c) inverted bow-tie, (d) inverted diabolo PAs for several values of the wing length $v$ of the antennas. }
  \end{center}
\end{figure}

\begin{figure}[h!]
  \begin{center}
    \includegraphics[width=0.9\linewidth]{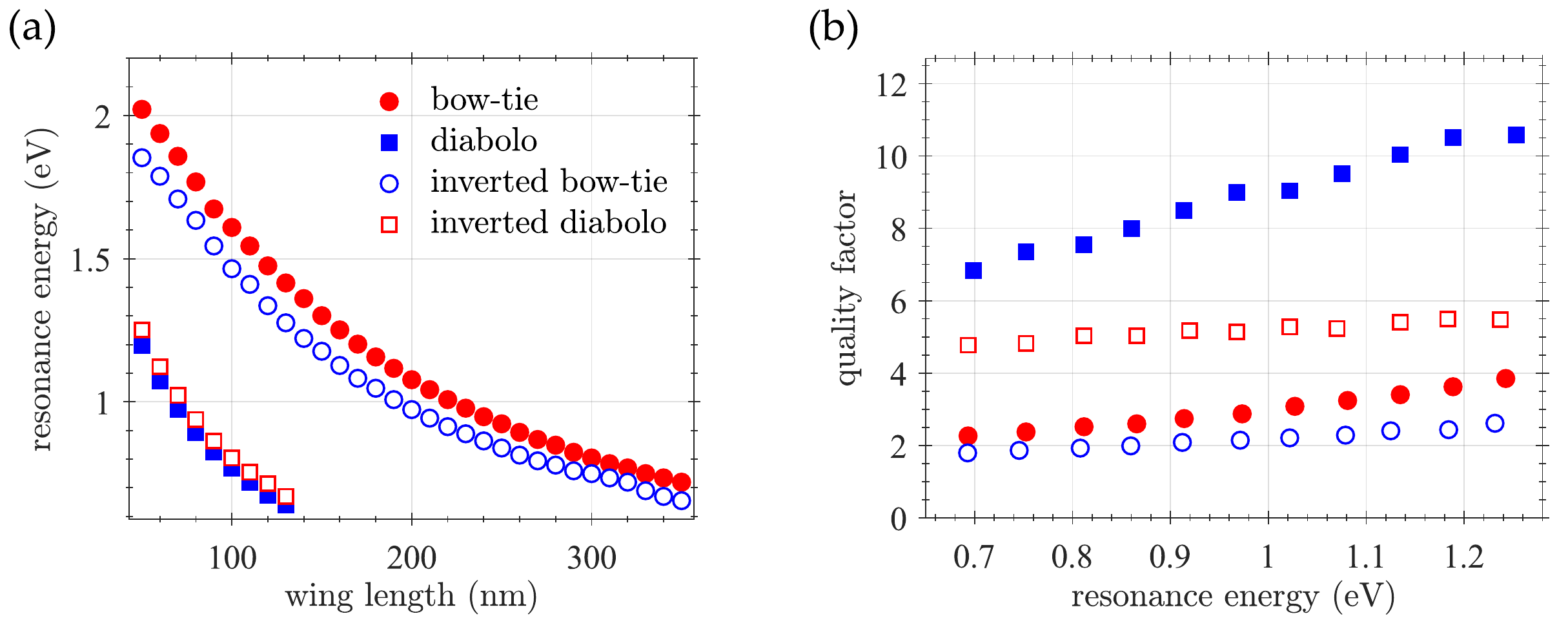}
    \caption{\label{figNO3} (a) Peak energies of the scattering efficiency $Q_\mathrm{scat}$ of bow-tie (red full circles), diabolo (blue full squares), inverted bow-tie (blue empty circles), and inverted diabolo (red empty squares) PAs as functions of the PA dimension -- length of the wing $v$. (b) Quality factors of LSP (represented by the peak energy of scattering cross-sections divided by their full width at half maximum). Notice that circles and squares correspond to bow-tie and diabolo PAs, full and empty symbols correspond to particles (direct PAs) and apertures (inverted PAs), and red and blue color corresponds to electric and magnetic hot spots, respectively. 
}
  \end{center}
\end{figure}

One of the quantities characterizing plasmonic response of antennas is their scattering efficiency $Q_\mathrm{scat}$. It describes the power $P_\mathrm{scat}$ scattered by the antenna illuminated with a monochromatic plane wave with an intensity $I_0$ and is defined as $Q_\mathrm{scat}=P_\mathrm{scat}/(I_0 \, S)$, where $S$ denotes the geometrical cross-section of the antenna. Spectral dependencies of $Q_\mathrm{scat}$ for all four PA types are shown in Fig.~\ref{figNO2} and the energies of the lowest scattering peak corresponding to a dipole plasmonic mode are shown in Fig.~\ref{figNO3}. We observe that Babinet's principle holds reasonably well. Peak energies of the scattering efficiency in the complementary PA (i.e., bow-tie and inverted bow-tie, diabolo and inverted diabolo) of the same size differ by less than 11\,\%. The difference is less pronounced for large antennas, in line with the requirements of Babinet's principle: perfectly thin and opaque metal.~\cite{Horak2018SciRep} For the bow-tie geometry, the scattering peaks of inverted PAs are less intense and red-shifted with respect to direct PAs (as is the case also for disc-shaped antennas~\cite{Horak2018SciRep}), while opposite is true for the diabolo geometry.

Not surprisingly, the peak energies of the scattering cross-section for the diabolo PAs are considerably smaller than that for the bow-tie PAs of the same wing length $v$. In other words, for the same energy, the diabolo PAs are smaller by a factor of more than 2 than the bow-tie PAs. This effect is explained by larger effective size of connected (i.e., diabolo) antennas in comparison with disjoint ones (bow-tie) and has been observed previously~\cite{doi:10.1021/nl304078v}. There is an important practical consequence. Bow-tie geometry allows to achieve high energies for which diabolo-type PAs can be too small for involved fabrication technique. Considering the minimum wing length of 50\,nm, diabolo antennas cover the LSP energy range up to 1.2\,eV while bow-tie antennas operate up to 2.0\,eV. On the other hand, diabolo geometry allows for a more compact PA design and better integration to more complex devices, such as a scanning near-field probe with the electric hot spot.~\cite{Mivelle:10}

Diabolo antennas, either direct or inverted, feature considerably narrower scattering peaks corresponding to larger quality factor than bow-tie antennas (Fig.~\ref{figNO3}). This is probably related to lower radiative losses due to their smaller volume.

\begin{table}[h!]
\begin{center}
\begin{tabular}{lcccc}
\hline
\hline
 & bow-tie & diabolo & inverted bow-tie & inverted diabolo \\
\hline
0.8~eV & 300\,nm & 95\,nm & 270\,nm & 100\,nm \\
1.8~eV & 75\,nm & - & 55\,nm & - \\
\hline
\hline
\end{tabular}
\caption{\label{table1}Dimensions (wing length) of the antennas featuring the lowest LSPR at the energy of 0.8~eV and 1.8~eV.}
\end{center}
\end{table}

In the following, we compare the properties and performance of all four types of PAs. We adjust the dimensions of the compared PAs so that they all feature the LSPR at the same energy. Table~\ref{table1} shows the dimensions of the antennas are listed in Table~\ref{table1} for two specific energies: 1.8\,eV corresponding to the minimum absorption of gold (i.e., minimum of the imaginary part of dielectric function) and 0.8\,eV corresponding to one of the optical communication wavelengths (1550\,nm). We note that the former energy is accessible only with bow-tie antennas. We have therefore focused at the energy of 0.8\,eV.


\begin{figure}[h!]
  \begin{center}
    \includegraphics[width=\linewidth]{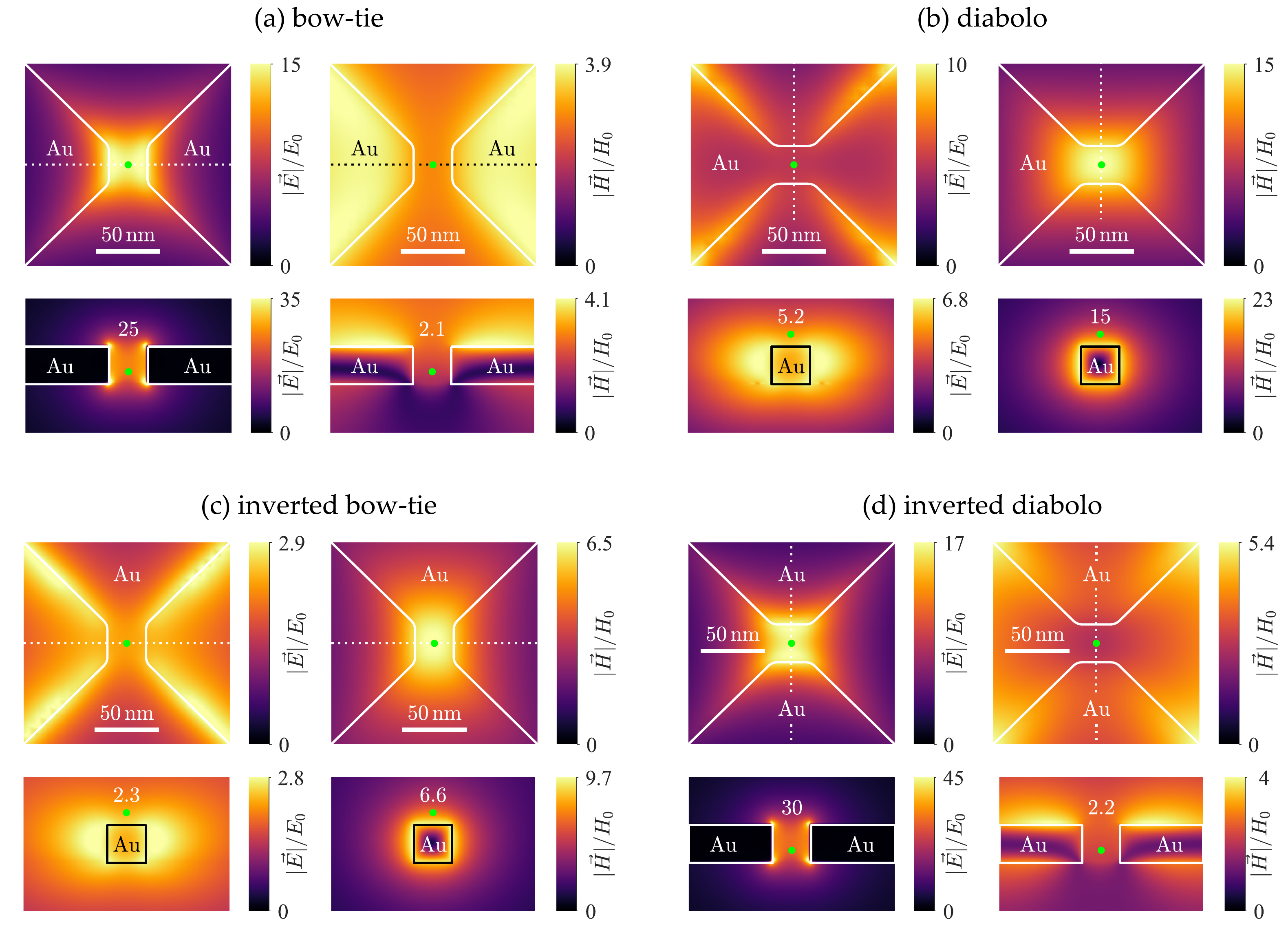}
    \caption{\label{figNO4} Planar cross-sections of the electric ($\vert \vec{E} \vert$) and magnetic ($\vert \vec{H} \vert$) field magnitudes divided by related magnitudes of the driving plane wave. The top two sub-plots in each panel show field distributions in the plane parallel to the PA plane, 10\,nm above the upper PA boundary. The bottom two sub-plots then show field distributions in the vertical plane with the orientation indicated by white dotted lines in the in-plane field plots. The size of all PAs was set so that they feature the lowest LSPR at 0.8\,eV, which is also the photon energy of the driving field. Solid white lines indicate antenna boundaries, while hot spots are marked by the green point and the numbers correspond to the field enhancement in the hotspot. The figures show only the central part of the antennas with the metallic parts being denoted as Au for clarity.}
  \end{center}
\end{figure}

Figure~\ref{figNO4} shows the formation of the hot spot. PAs featuring the lowest LSPR at the energy of 0.8\,eV are illuminated by a plane wave with the same photon energy. Bow-tie and inverted diabolo PAs feature the electric hot spot and delocalized magnetic field, while diabolo and inverted bow-tie PAs feature the magnetic hot spot and delocalized electric field. Interestingly, the volume of all the hot spots is comparable despite pronounced differences in the dimensions of PAs. The fields exhibit clear Babinet's duality: their spatial distribution in direct and complementary antennas is qualitatively similar with interchanged electric and magnetic components. Nevertheless, the magnitudes of the complementary fields differ rather significantly. As an example, the electric field within the hot spot of the bow-tie antenna has the relative magnitude around 25 while the magnetic field of the inverted bow-tie has maximal relative magnitude less than 10, i.e., almost three times weaker than expected. This observation is attributed to the finite thickness and conductivity of gold, which both limit the validity of Babinet's principle. As for the direct and inverted diabolo, the difference in the magnitudes of the electric and magnetic fields is less pronounced, but still quite significant. The bow-tie/diabolo duality can be observed for the field forming the hot spot (e.g., electric for bow-tie and magnetic for diabolo) which has very similar spatial distribution in both cases. However, the distribution of the delocalized field (e.g., electric for bow-tie and magnetic for diabolo) differs. In general, magnetic fields are weaker than electric fields since the energy of the electric field is only partly converted to the energy of magnetic field and the other part converts into kinetic energy of electrons~\cite{PhysRevLett.95.223902,Horak2018SciRep}.

\subsection*{Figures of merit for optical spectroscopy}


Plasmonic antennas can be used to enhance absorption and emission of light. Consequently, they enhance the signal of interest inthe signal of interest in various optical spectroscopy techniques, including absorption spectroscopy, Raman spectroscopy, photoluminescence spectroscopy, and absorption spectroscopy of magnetic transitions. Here we define figures of merit (FoM) for plasmonic enhancement of different spectroscopy techniques and evaluate them for all four type of PAs. We will consider a small volume of the analyzed material (e.g. molecule, quantum dot, nanostructured material, or just nanosized crystal) that fits into the size of the hot spot. 

In case of absorption spectroscopy, absorbed power can be expressed by Fermi's golden rule as $P=0.5\Re [\sigma(\omega)] |E|^2$ where $\omega$ is the frequency of the probing radiation (in the following referred to as light), 
$\sigma$ is the complex conductivity of the analyte, and $|E|$ is the magnitude of the electric component of light. For simplicity we consider both the electric field and transition dipole moment to be polarized along the axis of the PA. The presence of plasmonic antennas alters the magnitude of electric field exciting the analyte. For the driving field (a plane wave) with the electric field intensity $E_0$, the electric intensity in the hot spot reads $E_\mathrm{HS}$. We define the electric field enhancement $Z_E=E_\mathrm{HS}/E_0$. Clearly, absorbed power is enhanced by the factor of $Z_E^2$, which is thus suitable FoM for plasmon enhanced absorption spectroscopy. Raman scattering is a two-photon process, where each of the subprocesses, i.e., absorption of the driving photon and re-emission of the inelastically scattered photons, is enhanced by $Z_E^2$ (spectral dependence of $Z_E$ can be neglected considering low relative energy shift in the Raman scattering and large energy width of plasmon resonances). Therefore, FoM for the plasmon enhanced Raman spectroscopy reads $Z_E^4$. 

Absorption spectroscopy of magnetic transitions is relevant for the study of rare earth ions in the visible~\cite{PhysRevLett.114.163903,PhysRevB.96.224430}. Electron paramagnetic resonance is in principle also absorption spectroscopy involving magnetic dipole transitions in the microwave spectral range. Absorbed power can be expressed as $P=0.5\omega \Im [\mu(\omega)] |H|^2$ where $\omega$ is the frequency of light, $\mu$ is the complex permeability of the analyte, and $|H|$ is the magnitude of the magnetic component of light. For the magnetic field enhancement $Z_H$ defined analogously to $Z_E$, the FoM for absorption spectroscopy of magnetic transition reads $Z_H^2$.

\begin{figure}[h!]
  \begin{center}
      \includegraphics[width=0.55\linewidth]{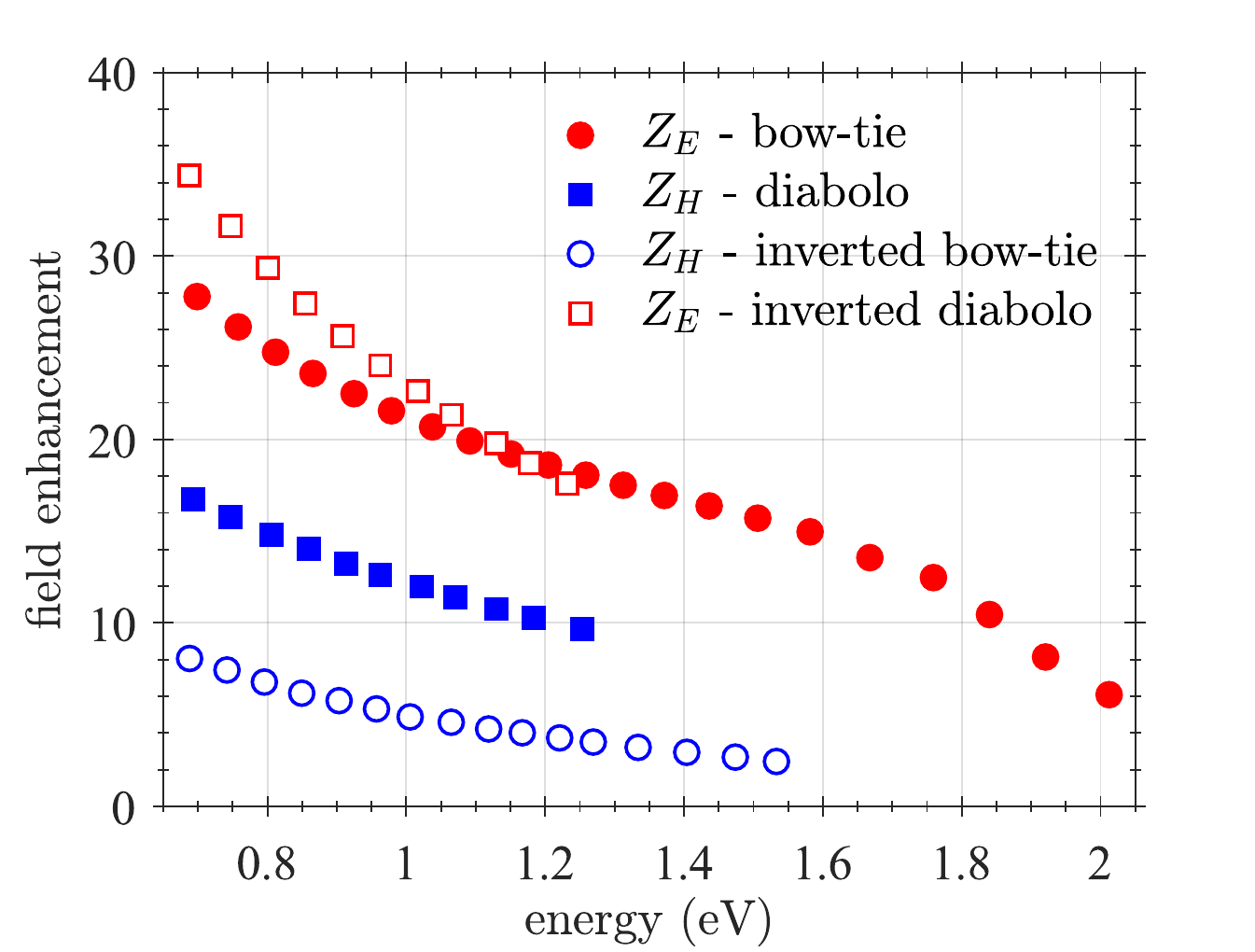}
    \caption{\label{figNO5}  Electric (red) and magnetic (blue) field enhancements for bow-tie, diabolo and their Babinet complements in their respective hotspots. The electric field enhancement $Z_E$, defined as the ratio of the local electric field intensity $\vert \vec{E}_{\mathrm{HS} } \vert$ and the amplitude of the driving field $E_0$, enters the figures of merit for plasmon-enhanced absorption spectroscopy ($\sim Z_E^2$) and Raman scattering ($\sim Z_E^4$). On the other hand, the magnetic field enhancement $Z_H$, defined as the ratio of the local magnetic field intensity $\vert \vec{E}_{\mathrm{HS} } \vert$ and the magnetic field intensity $H_0$ of the driving field, is important for plasmon-enhanced absorption spectroscopy of magnetic transitions ($\sim Z_H^2$).        
\del {(a) Figures of merit for plasmon-enhanced absorption spectroscopy (squares) and Raman scattering (circles), defined as $Z_E^2$ and $Z_E^4$, respectively. Electric field enhancement $Z_E$ is defined as the ratio of the electric field intensity magnitude $E_\mathrm{HS}$ in the hot spot and the electric field intensity magnitude $E_0$ of the driving field. (b) Figure of merit for plasmon-enhanced electron paramagnetic resonance, defined as $Z_H^2$. Magnetic field enhancement $Z_E$ is defined as the ratio of the magnetic field intensity magnitude $H_\mathrm{HS}$ in the hot spot and the magnetic field intensity magnitude $H_0$ of the driving field.\cc{Field enhancement or figures of merit?}\cc{Extend the range?}}
 }
  \end{center}
\end{figure}

We should note that the choice of the location in which we should evaluate the enhancement factors is somewhat arbitrary. In the case of the bow-tie and inverted diabolo, we decided to take the values from a spot positioned in the center of the gap, 10\,nm above the substrate, while for the inverted bow-tie and diabolo, the spot is situated 10\,nm above the center of the bridge. This choice gives us reasonable estimates that are close to the average values over the whole hot spots and it also ensures sufficient separation from the metal which is relevant for dipolar emitters and their quenching. Ultimately, we can afford this slight inconsistency in the definition of the hot spot as we always compare bow-tie with inverted diabolo and inverted bow-tie with diabolo, i.e. PAs with the same definition of the hotspot. With this in mind, we can turn our attention back to the field enhancements $Z_E$ and $Z_H$. The inspection of Figure~\ref{figNO5} shows that both $Z_E$ and $Z_H$ decrease with increasing energy as a consequence of decreased funneling effect (the size of the wings decreases while the size of the bridge or the gap is kept constant). The electric field enhancement $Z_E$ ranges between 18 and 34 with the inverted diabolo PA having slightly better performance than bow-tie. The magnetic field enhancement ranges between 10 and 17 for the diabolo PA but only between 4 and 8 for inverted bow-tie. Thus, inverted diabolo presents an excellent option for electric field enhancement while inverted bow-tie does not perform particularly well for the magnetic field enhancement.


Luminescence spectroscopy is another important method that can benefit from plasmon enhancement. We will consider a simple model based on the rate equation model. A metastable excitonic state with the degeneracy $g$ is populated through external excitation with the rate $\gamma_G$. The generation is only efficient when the metastable state is unoccupied. For its population $n$, the total generation rate reads $(g-n)\cdot \gamma_\mathrm{G}$. Excitons decay into the vacuum state via radiative and non-radiative recombination paths with the rates $\gamma_\mathrm{R0}$ and $\gamma_\mathrm{NR0}$, respectively. The rate equation reads
$$
\frac{dn}{dt}=(g-n)\cdot \gamma_\mathrm{G}-n\cdot \gamma_\mathrm{R0}-(g-n)\cdot \gamma_\mathrm{NR0}.
$$ 
In steady state,
$dn/dt=0$ and 
$$n=g\cdot \gamma_\mathrm{G} / ( \gamma_\mathrm{G} + \gamma_\mathrm{R0} + \gamma_\mathrm{NR0}).$$
Two regimes can be distinguished. In the linear (weak pumping) regime, $\gamma_\mathrm{G} \ll \gamma_\mathrm{R0} + \gamma_\mathrm{NR0}$ and 
$$n\approx g \gamma_\mathrm{G} / ( \gamma_\mathrm{R0} + \gamma_\mathrm{NR0}),$$
i.e., population is proportional to pumping. In the saturation (strong pumping) regime, $\gamma_\mathrm{G} \gg \gamma_\mathrm{R0} + \gamma_\mathrm{NR0}$) and $n\approx g$, i.e., metastable state is fully occupied. Emitted power reads 
$$P_\mathrm{PL}=n\cdot \gamma_\mathrm{R0} \cdot \hbar\omega,$$
where $\hbar\omega$ is the photon energy. In the linear regime, the emitted power can be expressed using internal quantum efficiency $\eta_0=\gamma_\mathrm{R0} / ( \gamma_\mathrm{R0} + \gamma_\mathrm{NR0})$ as
$$P_l=g \gamma_\mathrm{G}\eta_0 \cdot \hbar\omega$$
and in the saturation regime
$$P_s=g \gamma_\mathrm{R0} \cdot \hbar\omega.$$

The presence of plasmonic antennas affects all three processes (generation, radiative decay, and non-radiative decay). The effect on generation varies from very important in the case of photoluminescence~\cite{Kinkhabwala2009} to negligible in the case of electroluminescence. In general, generation is sequential inelastic process and cannot be described by a simple model. For that, we will not consider plasmon enhancement of generation in the following and focus on its influence of the radiative and non-radiative decay rates. 

Spontaneous emission is affected via Purcell effect~\cite{PhysRev.69.681}. The emitter transfers its energy to PA where it is partially radiated into far field and partially dissipated. It is customary to express the rates of both processes in multiples of the spontaneous emission rate $\gamma_\mathrm{R0}$: $Z_\mathrm{R}$ being the radiative enhancement and $Z_\mathrm{NR}$ the non-radiative enhancement~\cite{Atwater2010,edes2016}. Total radiative and non-radiative decay rates in the presence of plasmonic particles read $\gamma_\mathrm{R}=Z_\mathrm{R}\cdot \gamma_\mathrm{R0}$ and $\gamma_\mathrm{NR}=Z_\mathrm{NR} \cdot \gamma_\mathrm{R0}+\gamma_\mathrm{NR0}$, respectively.

The figure of merit for plasmon enhanced luminescence (only its emission part) is the rate of the powers emitted with and without the presence of the PA. For linear regime, FoM is 
$$F_l=\eta/\eta_0$$
while for the saturation regime it reads simply
$$F_s=Z_\mathrm{R}.$$
Consequently, only emitters with poor internal quantum efficiency can benefit from plasmon enhancement in the linear regime while the emitters with high internal quantum efficiency will suffer from the dissipation in metallic PA. On the other hand, in the saturation regime plasmon enhancement is benefitable as long as $Z_\mathrm{R}>1$.

\begin{figure}[h!]
  \begin{center}
      \includegraphics[width=\linewidth]{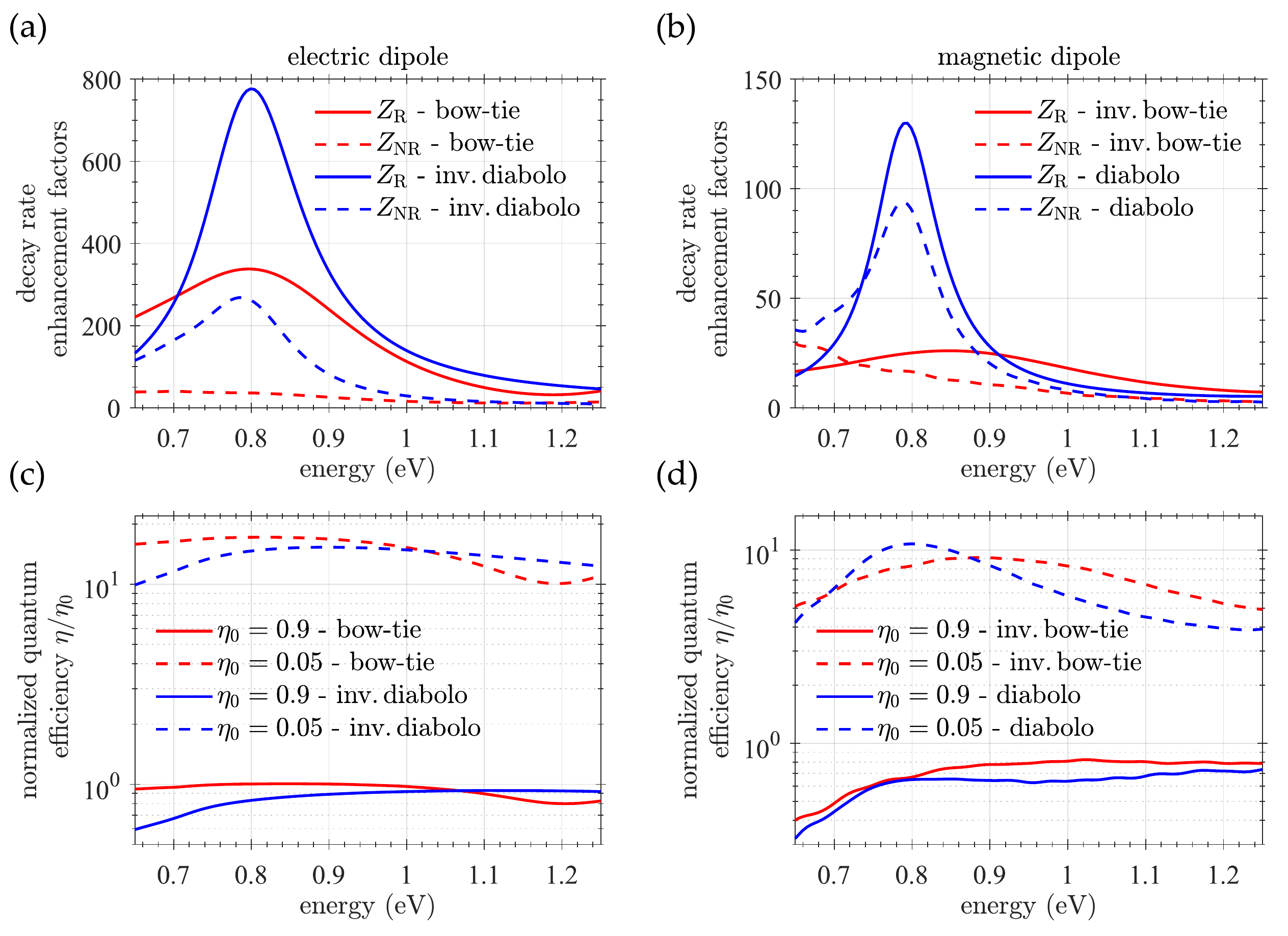}
    \caption{\label{figNO6} Radiative and non-radiative enhancement factors $Z_\mathrm{R}$ and $Z_\mathrm{NR}$, respectively, as functions of the photon energy for the (a) electric dipole transition and (b) magnetic dipole transition. The values are averaged over all possible polarizations of the transitions. Panels (c) and (d) then show enhancement of the overall quantum efficiency for two values of the internal quantum efficiency, namely $\eta_{0}=0.9$ (good emitter) and $\eta_{0}=0.05$ (poor emitter). Note that even though the radiative enhancement for direct and inverted diabolos is significantly higher than for their bow-tie counterparts, their enhancement of quantum effieciency is due to their equally larger non-radiative enhancement more or less the same.}
  \end{center}
\end{figure}

Figure~\ref{figNO6} shows spectral dependence of radiative and non-radiative enhancement factors ($Z_\mathrm{R}$ and $Z_\mathrm{NR}$, respectively) for different types of PA with the maximum field enhancement at 0.8\,eV. The point-like isotropic emitter (i.e., all polarizations are involved with the same intensity) was positioned to the centre of the PA 10~\,m above the substrate (bow-tie and inverted diabolo) or 10\,nm above the surface of gold (diabolo and inverted bow-tie). Such a separation shall suppress emission quenching due to non-radiative decay of the emitter. For the electric dipole transitions, large radiative enhancement (several hundreds) is obtained for both bow-tie and inverted diabolo. The inverted diabolo offers approximately twice larger peak enhancement. However, bow-tie benefits from much lower non-radiative enhancement and is thus preferable for most emitters in the linear regime. For magnetic dipole transitions, diabolo provides considerably larger radiative enhancement than inverted bow-tie, but it also suffers from the considerably larger non-radiative enhancement. In addition, the resonance of the inverted bow-tie is considerably wider which can prioritize this type of antenna for emitters with broad spectral bands. The preferred PA type therefore depends on specific application. We note that the peaks in the enhancement are spectrally shifted from the maximum field enhancement; the effect is particularly pronounced for the inverted bow-tie.

\subsection*{Babinet dimer with electromagnetic hot spot}

\begin{figure}[h!]
  \begin{center}
    \includegraphics[width=1\linewidth]{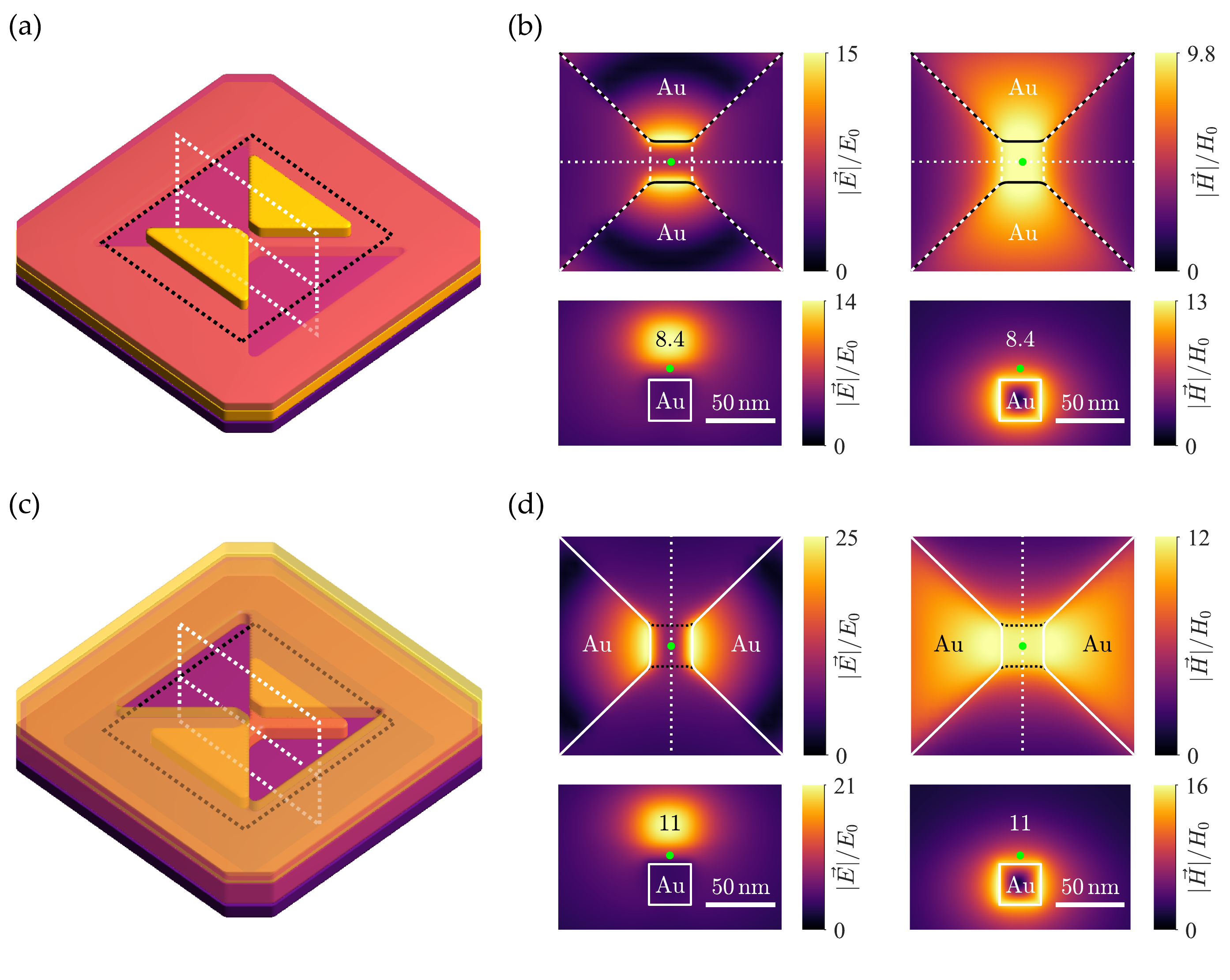}
    \caption{\label{figNO7} (a) Schematical drawing of the Babinet bow-tie dimer (BBD). A direct bow-tie (wing length 110\,nm) lies on top of an inverted bow-tie (wing length 200\,nm), they are mutually rotated by 90 degrees and separated by a 10\,nm spacer layer. (b) Distribution of the electric (left) and magnetic (right) fields in the vicinity of the BBD. The top two subplots show the fields in a plane parallel to the individual PAs, namely inside the spacer layer with 5\,nm distance from both of them [indicated by a black dotted line in panel (a)]. The bottom two subplots then show the fields in a vertical plane perpendicular to the metal bridge of the inverted bow-tie. To avoid any confusion regarding its orientation, it is outlined by a white dotted line in the schematical drawing in panel (a) and also in the top two subplots showing the fields in the horizontal plane. (c) Schematical drawing of the Babinet diabolo dimer (BDD). A direct diabolo (wing length 110\,nm) lies underneath an inverted bow-tie (wing length 200\,nm), they are mutually rotated by 90 degrees and separated by a 10\,nm spacer layer. The distribution of the electric (left) and the magnetic (right) fields around the BBD is plotted in (d), with planar cross-sections positioned and oriented in the same manner as in the case of BBD. Note that both BBD and BDD were illuminated with a plane wave polarized along the long axes of the direct PAs and the green point marks the position, in which the electric and magnetic field enhancements are equal in magnitude (with value specified by the number).}
  \end{center}
\end{figure}

Bow-tie and diabolo PAs enhance either electric or magnetic component of the field, while the other component is only weakly enhanced and spatially focused. In this section we propose Babinet dimer antenna that forms an electromagnetic hot spot enhancing and focusing both components of the electromagnetic field equally. The Babinet dimer antenna is formed by a direct and an inverse PA, vertically stacked so closely that their individual electric and magnetic hot spots overlap. We explore and compare two configurations, namely the Babinet bow-tie dimer (BBD) [schematically depicted in Figure~\ref{figNO7} (a)], consisting of a bow-tie on top of an inverted bow-tie PA, and the Babinet diabolo dimer (BDD) [sketched in Figure~\ref{figNO7} (c)] made up by an inverted diabolo on top of a diabolo PA. In both configurations, the upper PA is rotated with respect to the bottom one by 90 degrees so that both of them can be excited by the same source polarization (oriented along the long axis of the direct PA) and the upper and bottom PAs are separated by a 10\,nm spacer layer with refractive index equal to 1.5. As the individual modes in the closely spaced PAs exhibit strong interaction, the dimensions of the dimer constituents have been adjusted so that the maximum field enhancement occurs at 0.8\,eV for both the electric and magnetic component. For the BBD, the wing lengths of the top ($\uparrow$) and bottom ($\downarrow$) PAs were set to $v_{\uparrow}=110$\,nm and $v_{\downarrow}=200$\,nm, while for the BDD, the optimal dimensions read $v_{\uparrow}=200$\,nm and $v_{\downarrow}=110$\,nm. Note that the antenna providing the magnetic enhancement is in both cases situated underneath the one with the electric enhancement so that the electromagnetic hot spot is directly accessible from the top.

\del{Bow-tie and diabolo PAs enhance either electric or magnetic component of the field, while the other component is only weakly enhanced and spatially focused. In this section we propose Babinet bow-tie dimer (BBD) antenna that forms an electromagnetic hot spot enhancing and focusing both components of the electromagnetic field on equal footing. BBD is formed by a direct and an inverse bow-tie PA, vertically stacked so closely that their individual electric and magnetic hot spots overlap. The axes of the PAs can be either perpendicular (crossed BBD) or parallel (inline BBD), as shown in Fig.~\ref{figNO7}. In the crossed BBD both electric mode in the direct bow-tie and the magnetic mode in the inverted bow-tie are driven by the electric field polarized along the same direction. In the inline BBD, the electric mode in the direct antenna and the magnetic mode in the indirect antenna are driven by the electric field of mutually perpendicular polarization. This allows to tune the intensity of excitation of individual modes and their mutual phase shift. Naturally, the individual modes in the closely spaced PAs exhibit strong interaction influencing their optical properties. The dimensions of both bow-ties in the dimer have been therefore slightly adjusted so that the maximum field enhancement occurs at 0.8~eV for both the electric and magnetic component. }

Figure~\ref{figNO7}(b) demonstrates the formation of the electromagnetic hot spot in the BBD. The dimer is illuminated by the field polarized along the long axis of the bow-tie and perpendicular to the long axis of inverted bow-tie, which results into formation of an electric hot spot (field enhancement 14) around the direct and a magnetic hot spot (field enhancement 13) around the inverted bow-tie. Closely spaced hot spots overlap, yielding maximum simultaneous enhancement of both fields close to 8.4 at the position indicated by the green point in Fig.~\ref{figNO7}. The inspection of Figures~\ref{figNO3}~and~\ref{figNO5} reveales that these values are similar to those obtained for single PAs of comparable size. This indicates that the enhancement mechanism based on the charge accumulation (or the current funneling) at the wing apices is rather robust and resistant to changes in surroundings of the PA. The other proposed design, BDD, possesses electromagnetic hotspot as well, with maximum simultaneous enhancement of 11, while the individual maxima read 21 (electric enhancement) and 16 (magnetic enhancement) [see Figure~\ref{figNO7}(d)]. These values are again close to those encountered in single PAs, despite the partial screening of the bottom diabolo by its upper counterpart. On the whole, the better performance of isolated diabolos (at least in terms of local field enhancement) imprints itself also into Babinet dimers.

So far we have altogether disregarded the vectorial nature of electromagnetic fields, which can be important in certain applications. In the designs proposed above, the electric and magnetic fields in the hot spot are perpendicular to each other, but one can achieve also other mutual orientations simply by rotating the vertically stacked antennas with respect to each other. Such control over the local polarization state of the light is quite valuable, especially when we consider the aforementioned robustness with both field amplitude and orientation tightly bound to the geometry of the PAs.

In comparison to previous proposals~\cite{Yu:13,C3NR05536A} and realizations~\cite{doi:10.1021/acsphotonics.6b00857} of plasmonic electromagnetic hot spots, our proposal brings two benefits. (i) It enhances both field on equal basis, i.e., with the same amplitude, resonance frequency, and lateral spatial distribution. (ii) It involves two isolated antennas which can be adjusted independently, allowing extended tunability of the hot spot.


\del{Inline BBD offers more versatility. For the polarization of the driving field parallel to both antenna axes only the electric hot spot is formed [Fig.~\ref{figNO9}(top)] while for the perpendicular polarization only the magnetic hot spot is formed [Fig.~\ref{figNO9}(top)]. It is possible to apply the driving field with such a polarization that both the hot spots are excited with the identical field enhancement. In the considered case, the polarization angle was \cc{xxx~$^\circ$} (with respect to the axes of the antennas) and the enhancement in the joint hot spot was \cc{xxx}.  }

\section*{Conclusion}

We have focused on the plasmonic antennas featuring electric, magnetic, and electromagnetic hot spots: bow-tie and inverted diabolo, diabolo and inverted bow-tie, and their dimers, respectively. We have combined two types of electric-magnetic complementarity: bow-tie/diabolo duality and Babinet's principle.

For a specific resonance frequency, diabolo antennas were significantly smaller than bow-tie antennas, and thus harder to fabricate but easier to integrate. For the minimum wing length of 50~nm, bow-ties covered energy range up to 2.0~eV while diabolos only up to 1.2~eV. Diabolo antennas also exhibit considerably narrower resonances related to higher Q factor as a consequence of lower scattering cross-section.

We have evaluated figures of merit for different methods of optical spectroscopy. One of the most important is the field enhancement in the hot spot, which was larger for the diabolo antennas (and also for the electric field). For the luminescence, the key figure of merit is the radiative and non-radiative decay enhancement. Here, diabolo antennas exhibited slightly stronger radiative decay enhancement but also pronouncedly stronger non-radiative enhancement, making bow-tie antenna a preferred option for the electric dipole transitions and inverted bow-tie and equivalent alternative of diabolo for the magnetic dipole transitions.

Finally, we have proposed Babinet dimer antennas enhancing both the electric and magnetic field on equal basis and forming electromagnetic hot spot, which finds applications in studies of rare earth ions, optical trapping, metamaterials, or non-linear optics.

\section*{Methods}

\subsection*{Simulations}
In all simulations, the bow-tie and diabolo antennas have been represented by two gold triangles or triangular apertures (as shown in Fig.~\ref{figNO1}) of 30\,nm height on a semiinfinite glass substrate. Babinet dimers are formed by two complementary PAs (direct and inverted, each of 30\,nm height) vertically separated by a 10\,nm thick layer with reractive index equal to 1.5. The whole dimer lies on a semiinfinite glass substrate. The dielectric function of gold was taken from Ref.~\cite{PhysRevB.6.4370} and the refractive index of the glass was set equal to 1.47.

The electromagnetic field has been calculated with finite-difference in time-domain (FDTD) method using a commercial software Lumerical.

Scattering efficiencies and the near-field distribution have been calculated using plane wave as an illumination. Transition decay rates have been calculated as the decay rate of the power radiated by oscillating electric or magnetic dipole into its surrounding (total decay rate) and into far field (radiative decay rate). The dipole has been positioned at the vertical symmetry axis of the antenna with polarization parallel with the polarization of the plasmonic near field. Its height above the antenna plane has been set to 10\,nm. \del{varied to find optimum figures of merit.}     

\section*{Data availability}
The datasets analysed during the current study are available from the corresponding author on reasonable request.

\bibliography{manuscript}

\section*{Acknowledgement} 
We acknowledge the support by the Czech Science Foundation
(grant No.~17-25799S), Ministry of Education, Youth and Sports of the Czech Republic
(projects CEITEC 2020, No.~LQ1601, and CEITEC Nano RI, No.~LM2015041), and
Brno University of Technology (grant No.~FSI/STI-J-18-5225).

\section*{Author information}
\subsection*{Contributions} 
V.K. conceived and coordinated research with help of T.S. Ma.H. and A.K. performed numerical simulations. All authors were involved in the data processing and interpretation. V.K. and Ma.H. wrote the manuscript.  

\subsection*{Competing interests}
The authors declare no competing interests.

\end{document}